\documentclass[useAMS,usenatbib]{mn2e}
\usepackage{psfig}

\title[Where do early-type void galaxies come from?]
{Where do ``red and dead'' early-type void galaxies come from?}  

\author[Croton \& Farrar]{
\parbox[t]{\textwidth}{
Darren J. Croton$^1$,
Glennys R.\ Farrar$^2$
}
\vspace*{6pt} \\ 
$^1$Department of Astronomy, University of California, Berkeley, CA, 94720, USA \\
$^2$Center for Cosmology and Particle Physics, Department of Physics,
  New York University, New York NY 10003
\vspace{-0.5cm} 
}

\date{Accepted ---. Received ---;in original form ---}

\pubyear{2005}

\newcommand{\plotone}[1]
           {\centering \leavevmode \psfig{file=#1,width=\columnwidth,clip=}}

\newcommand{\plotscaled}[1]
           {\centering \leavevmode \psfig{file=#1,width=16.0cm,clip=}}

\def\simlt{\lower.5ex\hbox{$\; \buildrel < \over \sim \;$}}
\def\simgt{\lower.5ex\hbox{$\; \buildrel > \over \sim \;$}}

\newcommand{\eightMP}{$8\,h^{-1} {\rm Mpc}$}

\begin{document}

\maketitle


\begin{abstract}

Void regions of the Universe offer a special environment for studying
cosmology and galaxy formation, which may expose weaknesses in our
understanding of these phenomena.  Although galaxies in voids are
observed to be predominately gas rich, star forming and blue, a
sub-population of bright red void galaxies can also be found, whose
star formation was shut down long ago.  Are the same processes that
quench star formation in denser regions of the Universe also at work
in voids?

We compare the luminosity function of void galaxies in the 2dF Galaxy
Redshift Survey, to
those from a galaxy formation model built on the Millennium
Simulation.  We show that a global star formation suppression
mechanism in the form of low luminosity ``radio mode'' AGN heating is
sufficient to reproduce the observed population of void
early-types. Radio mode heating is environment independent other than
its dependence on dark matter halo mass, where, above a critical mass
threshold of approximately $M_{\rm vir}\!\sim\!10^{12.5} M_\odot$, gas
cooling onto the galaxy is suppressed and star formation subsequently
fades.  In the Millennium Simulation, the void halo mass function is
shifted with respect to denser environments, but still maintains a
high mass tail above this critical threshold.  In such void halos,
radio mode heating remains efficient and red galaxies are found;
collectively these galaxies match the observed space density without
any modification to the model.  Consequently, galaxies living in
vastly different large-scale environments but hosted by halos of
similar mass are predicted to have similar properties, consistent with
observations.

\end{abstract}


\section{Introduction}
\label{sec:intro}

According to conventional cosmology, voids provide a special
laboratory for studying galaxy formation, in which the dark matter
density is locally that of a low density Universe.  Peebles has
claimed \citep{Peebles2001} that $\Lambda$ cold dark matter
($\Lambda$CDM) simulations overproduce void galaxies in comparison to
observation and he suggested that this may be evidence of
as-yet-undiscovered fundamental physics not included in $\Lambda$CDM
cosmology.  Previous work has laid the groundwork for examining this
question quantitatively \citep[e.g.][]{Mathis2002, Benson2003b,
Mo2004, Croton2004, Croton2005, Hoyle2004, Hoyle2005, Rojas2004,
Rojas2005, Colberg2005, Goldberg2005, Clemens2006, Xia2006,
Patiri2006, Sorrentino2006, vonBendaBeckmann2008}.  In
\cite{Croton2005} we introduced a practical definition of the void
luminosity function and measured it for early- and late-type galaxies
in the 2dF Galaxy Redshift Survey (2dFGRS).  Here, we investigate
whether $\Lambda$CDM simulations can describe the observed void
luminosity functions quantitatively, or whether new physics may be
required.

State-of-the-art modelling is not sufficiently advanced to treat all
the physical processes of galaxy formation from first principles.
Here we adopt a semi-analytic approach \citep{White1991} in which the
Millennium Simulation of dark matter is coupled with an analytic
simulation for baryonic evolution in a cosmologically evolving
background.  A simplifying approximation made in the current
generation of semi-analytic models is that the process of galaxy
formation depends only indirectly on the environment or age of a given
dark matter halo, via their impact on the halo mass function.  While
this clearly cannot be exactly true \citep[e.g.][]{Croton2007}, it
is interesting to ask whether such simplified models are capable of
reproducing the observed luminosity distributions of early- and
late-type galaxies in voids within the observational uncertainties.
As we show in the following, the answer is affirmative.

Key to the success of semi-analytic models in describing the global
galaxy luminosity function is the presence of heating mechanisms to
shut off star formation \citep{Benson2003}.  The exact processes
operating are currently subject to some debate.  What is not in
dispute is the observational evidence of galaxies at the bright-end of
the galaxy luminosity function with ongoing, low-luminosity active
nuclei. It is thus no surprise that most current semi-analytic models
incorporate active galactic nuclei (AGN) heating as a key ingredient.
\cite{Croton2006} called this the ``radio mode'' of AGN evolution.
Another process which can shut off star formation in subhalos is the
stripping of gas when a satellite falls into a larger halo, called
``strangulation''.

There is clear evidence of a population of ``red and dead'' early-type
galaxies\footnote{The term ``red and dead'' refers to those galaxies
on the red sequence which have had no active star formation in the
last several Gyr} even in voids \citep{Croton2005}, yet it is not
obvious whether either or both of these mechanisms (i.e. AGN heating
and strangulation) can operate efficiently in voids or whether some
other mechanism is needed.  We investigate this question via
semi-analytic simulations, and find that strangulation plays a minor
role.  We find that heating when the central galaxy passes the
critical mass is sufficient to account for the observed abundance of
early-type galaxies in voids.  The different luminosity functions of
early-type galaxies between voids and mean density environments can
thus be entirely attributed to the difference in the halo mass
function in the two environments, within the accuracy of present
measurements.  This explanation for the early-type luminosity function
has observable consequences as discussed below.  If these are
verified, it will lend credence to the physical interpretation of the
semi-analytic model.

This paper is organised as follows. Sections~\ref{sec:model} and
\ref{sec:density} describe the galaxy formation model and our measure
of environment within the Millennium Simulation, respectively.  We
present our void galaxy analysis in Section~\ref{sec:results} and
discuss our results in light of recent work and also within the
broader context of galaxy formation theory.  Finally,
Section~\ref{sec:summary} provides a brief summary.  Throughout we
assume a standard WMAP first year $\Lambda$CDM cosmology
\citep{Spergel2003, Seljak2005} and Hubble parameter
$H_0=100\,h^{-1}$kms$^{-1}$Mpc$^{-1}$.

\section{The Galaxy Formation Model}
\label{sec:model}

The galaxy formation model we use to study void environments is
identical to that described in \cite{Croton2006} (including parameter
choices). This model is implemented on top of the Millennium Run
$\Lambda$CDM dark matter simulation \citep{Springel2005}.  Below we
briefly outline the relevant aspects of the simulation and model to
our current work, and refer the interested reader to the above
references for further information.

The Millennium Run simulation follows the dynamical evolution of
$10^{10}$ dark matter particles in a periodic box of side-length
$500\,h^{-1}$Mpc with a mass resolution per particle of $8.6\times
10^8\,h^{-1}{\rm M}_{\odot}$.  It adopts cosmological parameter values
consistent with a combined analysis of the 2dFGRS \citep{Colless2001}
and first year WMAP data \citep{Spergel2003, Seljak2005}.
Friends-of-friends (FOF) halos are identified in the simulation using
a linking length of 0.2 the mean particle separation, while
substructure \emph{within} each FOF halo is found with an improved and
extended version of the {\small SUBFIND} algorithm of
\citet{Springel2001}.  Having determined all halos and subhalos at all
output snapshots, the hierarchical merging trees are constructed; these
describe in detail how structures grow as the universe evolves.  These
trees form the backbone onto which we couple our model of galaxy
formation.

Inside each tree, virialised dark matter halos at each redshift are
assumed to attract ambient gas from the surrounding medium, from which
galaxies form and evolve.  Our model effectively tracks a wide range
of galaxy formation physics in each halo using simple parametrised
forms, including reionization of the inter-galactic medium at high
redshift, radiative cooling of hot gas and the formation of cooling
flows, star formation in the cold disk and the resulting supernova
feedback, black hole growth and AGN feedback through the `quasar' and
`radio' epochs of AGN evolution, metal enrichment of the
inter-galactic and intra-cluster medium, and galaxy morphology shaped
through mergers and merger-induced starbursts.  At $z\!=\!0$ the
galaxy formation model contains approximately 9 million galaxies
brighter than our completeness limit of $M_{\rm b_J}\!-\!5\log_{10}\!
h\!=\!-15.8$ (representing the mean luminosity of a central galaxy in
a halo containing 64 particles).

A critical feature for the success of this model is the inclusion of
an AGN component and the separation of AGN into their high and low
accretion states, called the ``quasar mode'' and ``radio mode''
respectively.  Within this picture the two AGN modes are distinct in
their cause and effect.  Importantly for the work of
\cite{Croton2006}, the radio mode was used to suppress the cooling of
gas onto central galaxies living in group and cluster-sized halos, and
is primarily important at late times (typically $z\!<\!1$) and not
earlier (where the SFR density of the universe peaks,
e.g. \citealt{Madau1996}).  This model was tuned to provide a good
match to the local luminosity function and colours of galaxies.  The
assumed radio mode black hole accretion rate,
\begin{equation}
\dot{m}_{\rm BH} \propto m_{\rm BH} \, f_{\rm hot} \, V_{\rm vir}^3~,
\end{equation}
is very efficient above a critical halo mass threshold which is
approximately constant with time.  This mass turns out to be $M_{\rm
vir} \approx 10^{12 - 13} M_{\odot}$.  From this point-of-view, halos
grow with time until they cross the critical halo mass, after which
the radio mode dominates, cooling gas slows, then stops, and star
formation shut-down follows.  Note that no environment dependence has
been assumed in this implementation of radio mode heating.  We will
return to this point in Section~\ref{sec:results}.

\section{Estimating local densities}
\label{sec:density}

To facilitate a fair comparison between the model and observed
luminosity functions we determine environments in the Millennium
Simulation box in the same way as \cite{Croton2005} did for the
2dFGRS.  This is actually much simpler to do than with a magnitude
limited redshift survey, as one has a perfect measure of survey
geometry (a box in our case with volume $V_{\rm box}$) and selection
function (uniform and complete).  We consider the redshift zero
snapshot of the model, which we call the ``local'' population.  Note
that, although the 2dFGRS has a median redshift of $z\!\sim\!0.1$, all
2dFGRS magnitudes are k-corrected to $z\!=\!0$, and for our purposes
we assume that the evolution of structure between $z\!=\!0.1$ and
$z\!=\!0$ is negligible.  All model galaxies are shifted into redshift
space using the distant observer approximation.

\begin{figure}
\plotone{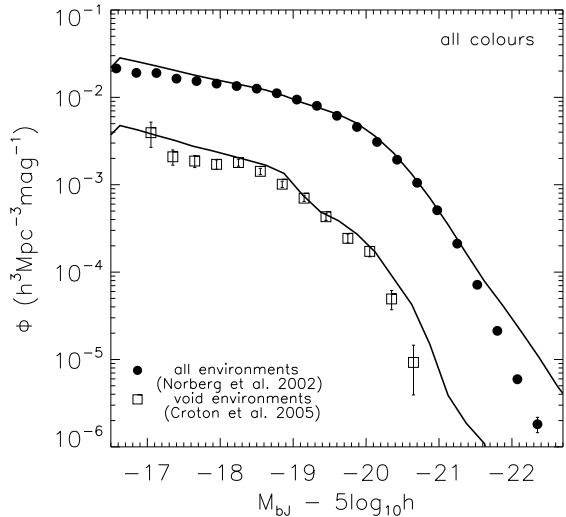}
\caption{The galaxy luminosity functions for all environments
\citep[filled circles,][]{Norberg2002} and in void regions exclusively
\citep[open squares,][]{Croton2005}, both from the 2dFGRS.  The
matching solid lines indicate the semi-analytic model result analysed
in the same way.  The model is tuned for good agreement with the
global 2dFGRS luminosity function; from this alone the distribution of
galaxy magnitudes in voids is also reproduced.}
\label{fig:LFall}
\end{figure}

\begin{figure}
\plotone{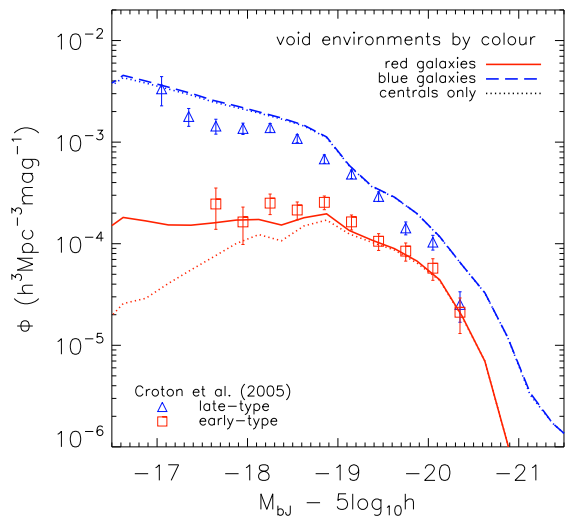}
\caption{The galaxy luminosity function in voids, now broken into
their blue/late-type and red/early-type sub-populations (see figure
legend).  The dotted lines show the model luminosity functions for
central galaxies only, i.e. when satellites are removed (see
Section~\ref{subsec:col}).  Both red and blue sequence model galaxies
have similar luminosity distributions to the 2dFGRS data.  The model
early-type void population is dominated by central galaxies.}
\label{fig:LFcol}
\end{figure}

Environments are measured using the smoothed top-hat number density
contrast around each galaxy.  As in \cite{Croton2005}, we define a
{\em density defining population} (DDP) of galaxies which fix the
density contours in the redshift space volume containing the model
galaxies; this is simply all galaxies in the magnitude range
$-19\!>\!M_{\rm b_J}\!-\!5\log_{10}\! h\!>\!-22$.  We then consider
\emph{all} galaxies in turn, counting the number of DDP neighbours
within a sphere of \eightMP\ radius, $N_{g}$.  This scale was found by
\cite{Croton2005} to optimise the simultaneous sampling of both the
under- and over-dense regions of the survey in a consistent way.  The
mean number density of the DDP is known, $\bar{\rho}$, and thus the
expected mean number within \eightMP\ is also known,
$\bar{N_{g}}\!=\!\bar{\rho}\,\frac{4}{3}\pi 8^3$.  Hence, the density
contrast, $\delta_8$, for each galaxy is simply
\begin{equation}\label{delta8}
\delta_8 \equiv \frac{\delta \rho_g}{\rho_g} =
\frac{N_g-\bar{N_g}}{\bar{N_g}}\ \Bigg{|}_{R=8h^{-1}\rm{Mpc}} ~.
\end{equation}
The Millennium Simulation was run using periodic boundary conditions,
and hence we do not need to worry about boundary effects in our
semi-analytic model galaxy catalogue or $\delta_8$ calculation.

To determine the volume fraction that voids occupy in the Millennium
Simulation we fill the simulation box with a large number of randomly
placed points, and for each we determine the number of DDP galaxies
within \eightMP.  This provides a uniform sampling of $\delta_8$ via
Equation~\ref{delta8} throughout the box.  The fraction of random
points in a $\delta_8$ range provides the volume filling fraction that
this environment occupies.  This allows us to appropriately normalise
the number density of galaxies in that environment when calculating
each luminosity function.  This method is described in detail in
Appendix~A of \cite{Croton2005}.

\section{Results and Discussion}
\label{sec:results}

\subsection{The void luminosity function}
\label{subsec:LF}

\begin{figure*}
\plotscaled{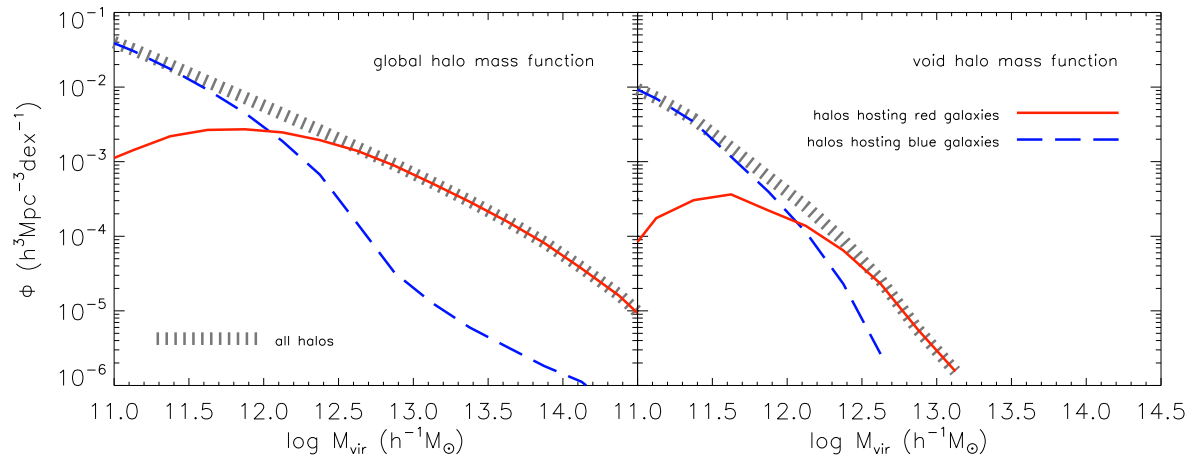}
\caption{The Millennium Simulation halo mass function (left panel) and
halo mass function in voids (right panel).  In both panels the halo
mass function is broken up into those who host red central galaxies
(solid lines) and those that host blue central galaxies (dashed
lines).  Red sequence galaxies occupy the most massive halos in all
environments -- these halos are subject to the radio mode low
luminosity AGN heating that ultimately shuts down star formation.
Notably, this shutdown begins at approximately the same mass in both
panels, $M_{\rm vir} \sim 10^{12-12.5} M_\odot$.  This implies that
the (environment independent) radio mode heating, plus a shift in the
halo mass function with environment, is sufficient to reproduce the
observed abundance of void early-type galaxies seen in
Figure~\ref{fig:LFcol}.}
\label{fig:MF}
\end{figure*}

We begin with Figure~\ref{fig:LFall}, which shows the luminosity
function of model galaxies for both the population as a whole, and all
galaxies in void regions.  Here we define a void as \cite{Croton2005}
did, $\delta_8\!<\!-0.75$, i.e. where the density within an \eightMP\
radius of the galaxy is less than $25\%$ the mean.  Over-plotted are
the observational results for the full 2dFGRS \citep[filled
circles,][]{Norberg2002} and void galaxy luminosity function
\citep[open squares,][]{Croton2005}.  

As was the focus of \cite{Croton2006}, the global luminosity
distribution of model galaxies is a good match to the local
observations.  Important model ingredients that produced this result
are the supernova in shallower potentials that expel disk gas to
reduce star formation and flatten the faint-end slope, and the
inclusion of an AGN heating source in large halos to starve massive
central galaxies of star forming fuel, resulting in the observed
bright-end exponential cut-off, as described in
Section~\ref{sec:model}.  Without these two critical aspects in the
model, the luminosity function instead would have a power-law shape
reflecting the underlying mass function of halos \citep{Benson2003}.

Note that there remain discrepancies in the top curve of
Figure~\ref{fig:LFall}, notably that the model over-predicts the
abundance of very bright galaxies.  As discussed in \cite{Croton2006},
these galaxies represent a population undergoing strong star formation
and merger-induced starbursts.  In such ultraluminous infrared
galaxies (ULIRGs), nearly all the light from young stars is absorbed
by dust and re-radiated in the mid- to far-infrared
\citep{Sanders1996}.  Improved modelling of the effects of dust are
required to adequately reproduce the properties of such systems.

The lower solid line in Figure~\ref{fig:LFall} shows the model result
for the void galaxy population in the Millennium Simulation.  The good
overall agreement is a result of the physical prescriptions and global
parameter choices in the \cite{Croton2006} model and not additional
fine-tuning (of which there was none).  There is some over-prediction
of the very brightest void galaxies.  However the discrepancy is not
significant enough to expect that improvements to existing aspects of
the model, e.g. dust as described above, cannot alleviate such
differences.  Additionally, we find a small over-prediction of faint
($M_{\rm b_J}\!-\!5\log_{10}\! h \simgt -18.5$) galaxies.
Unfortunately, systematic variations in the observed faint-end galaxy
luminosity function exist at a level greater than this (e.g. contrast
\citealt{Cole2001} to \citealt{Huang2003}).

Figure~\ref{fig:LFall} alone answers the challenge posed by Peebles
and outlined in the Introduction.  Specifically, in a $\Lambda$CDM
Universe the distribution of dark matter halos in voids, coupled with
a realistic model of galaxy formation, is consistent with voids as
observed in the real Universe.  Such regions are typically not devoid
of galaxies as claimed in \cite{Peebles2001}.

The remainder of this letter will aim to dissect and understand the
good agreement shown in Figure~\ref{fig:LFall} between model and
observation.

\subsection{Galaxy colours in voids}
\label{subsec:col}

The colour of a galaxy tells us a lot about the relevant physics that
has been dominant during its evolution.  A red spectrum typically
indicates that a star formation shutdown mechanism has been operating
for a significant part of the galaxy's lifetime.  We can use this
knowledge, in conjunction with our theoretical model, to gain insight
into how shutdown may occur and to what degree it may (or may not) be
environment specific.

To this end, in Figure~\ref{fig:LFcol} we again plot the void galaxy
luminosity function for both the 2dFGRS \cite[symbols,][]{Croton2005}
and semi-analytic model (lines), but now broken up by galaxy spectral
type (early/late) or colour (red/blue).  Early-type (triangles) and
late-type (squares) 2dFGRS galaxies are determined using the principal
component analysis of \cite{Madgwick2003}.  For the model we use the
bi-modal galaxy colours to separate red (solid line) from blue (dashed
line) at $m_{b_{\rm J}}\!-\!m_{r_{\rm F}}\!=\!1.07$ \citep[see
figure~9 of][]{Croton2006}.  Comparing model to observation,
Figure~\ref{fig:LFcol} shows agreement (to within $2\sigma$) between
the two sets of early-type/red and late-type/blue void luminosity
functions.  This has occurred as a byproduct of matching the model to
the observed \emph{global} properties and only these -- no special
void environment physics was needed.  Adding detail to the physical
prescriptions assumed by the model would presumably improve the
agreement further, but this is not our focus here.

The early-type 2dFGRS void galaxies in Figure~\ref{fig:LFcol} return
our focus to the question asked in the Introduction: how does one
understand early-type galaxies in voids when void environments are
typically very gas rich and slowly evolving, which tends to promote
star formation rather than suppress it.  At this point it may be
valuable to backtrack somewhat and revisit the two main mechanisms
acting in the model that can turn blue galaxies red. The first is
important for satellite (i.e. non-central) galaxies.  Upon infall into
a more massive system, any extended hot gas around the (now) satellite
is stripped by the denser medium and added to the more massive halo.
Such ``strangulation'' drives a satellite galaxy to redden rapidly
once its remaining cold disk gas is exhausted.  The second is the
radio mode heating discussed above and in Section~\ref{sec:model}.
Radio mode AGN operate only in central galaxies where the host halo
has grown above a critical mass threshold, approximately $M_{\rm vir}
\sim 10^{12.5} M_\odot$.

So what physical processes have occurred to produce in the observed
2dFGRS early-type void population?  It may be that such red galaxies
are simply a population of satellites and have survived long enough
for an effect on their colours to be seen.  To test this idea using
the model we perform a simple and revealing exercise.  From our
knowledge of which galaxies are central and which are satellites, we
remove those that are satellites and hence those that could have
experienced the strangulation effect.  This is shown by the dotted
lines in Figure~\ref{fig:LFcol}.  Although satellites do contribute to
the void early-type luminosity function at $M_{\rm
b_J}\!-\!5\log_{10}\! h \simgt -18.5$, they are clearly a minor
component brighter than this.  Indeed, all bright early-type void
galaxies are centrals in our galaxy formation model.

The second of our star formation shut-down mechanisms, i.e. radio mode
low luminosity AGN heating, requires that central galaxies reside in
sufficiently large dark mater halos.  Such halos are not expected to
be common in void regions of the Universe where low mass halos
dominate.  In low mass halos the central density of hot gas is not
high enough to maintain the low Eddington accretion needed to power a
central AGN outflow.  We check this explicitly in Figure~\ref{fig:MF},
where we plot the halo mass function for halos in all environments
(left panel) and void environments (right panel) (wide-dashed lines in
both).  In addition, we break the halo mass functions into those halos
hosting red central galaxies (solid lines) and those hosting blue
central galaxies (dashed lines).

Figure~\ref{fig:MF} reveals the origin of the void early-type
population -- they are simply central galaxies that live in halos
massive enough to have the low luminosity radio mode operating.  In
other words, the physics assumed by the model that transforms blue
galaxies into red operates independent of environment but can still
produce the observed environmental trends in voids (note that the
transition from blue-dominated to red-dominated dark matter halos
begins at approximately the same mass in both panels, $M_{\rm vir}
\sim 10^{12} M_\odot$).  Specifically, \emph{it is the shift in halo
mass function with environment that changes, not the transformation
mechanism itself.}

This picture has important consequences for galaxy evolution more
generally.  First, it indicates the necessity for active galaxies in
voids, confirmed in a number of studies (see, e.g.,
\citealt{Constantin2008} at $z\sim0$, and \citealt{MonteroDorta2008}
at $z\sim1$).  Second, it requires that void galaxies should have
similar properties to those in all other environments, assuming the
galaxies compared are hosted by halos of similar mass.  This is a
theoretical confirmation of numerous observational studies that have
shown that environments on scales larger than a few Mpc (i.e. outside
the dark matter halo) are not important for galaxy formation
\citep[e.g. see][and references therein]{Blanton2007}. For example,
\cite{Patiri2006} investigate the properties of void galaxies in the
Sloan Digital Sky Survey and compare to those of a similarly constructed mock catalogue
using the same \cite{Croton2006} model used here.  They found little
difference in the mean properties of void galaxies relative to the
field, specifically in colour, specific star formation rate, and
morphology.  Theoretically one may expect large-scale environment to
be important, e.g. the effect of assembly bias for dark matter halos
\citep{Gao2005}.  However the galaxy population does not appear to be
overly sensitive to this \citep[see, for example,][]{Croton2007}.  A
similar conclusion was recently arrived at in a complimentary analysis
by \cite{Tinker2008}.

\section{Summary}
\label{sec:summary}

It has been suggested that void environments pose a problem for galaxy
formation theory in a $\Lambda$CDM universe due to the apparent
over-abundance of dark matter halos in voids relative to the observed
abundance of galaxies \citep{Peebles2001}.  In this paper we show that
no conflict exists within the current observational uncertainty.

\begin{itemize}
\item Our model can reproduce the observed abundance of void galaxies,
both globally and by colour, without any modification or additional
environment dependent physics.
\item Early-type ``red and dead'' red sequence galaxies appear
naturally in the voids.  They arises because of a shift in the halo
mass function in low density environments combined with an environment
independent star formation shut-down mechanism efficient above a
critical halo mass (here, radio mode AGN).  Together, these
approximately produce the correct observed abundance.
\item Some notable consequences follow from our results.  For example,
at a given host halo mass, void galaxies are expected to have similar
properties on average to those in the field, because such galaxies will have had
similar evolutionary histories to field and cluster galaxies.
\end{itemize}

Voids and void galaxies provide an important probe of both cosmology
and galaxy formation.  Future wide-field surveys will produce
large-scale maps of the Universe out to high redshift, from which void
evolution can be studied in detail.  Such work will further constrain
galaxy formation theory, both in under-dense and the wider field
environment.

\section*{Acknowledgements}

DC acknowledges support from NSF grant AST00-71048, and wishes to
thank the NYU Center for Cosmology and Particle Physics for their
hospitality during part of this work.  GRF acknowledges NSF grants
NSF-PHY-0401232 and NSF-PHY-0701451.  The Millennium Simulation was
carried out by the Virgo Supercomputing Consortium at the Computing
Centre of the Max-Planck Society in Garching. The semi-analytic galaxy
catalogue used in this paper is publicly available at
http://www.mpa-garching.mpg.de/Millennium.

\bibliographystyle{mnras}
\bibliography{../paper}

\label{lastpage}

\end{document}